\documentclass[12pt]{article}

\usepackage{amsmath,amsthm,amsfonts,amssymb,amscd, latexsym}
\usepackage[dvips]{graphicx}

\begin{document}

\theoremstyle{plain}
\newtheorem{thm}{Theorem}
\newtheorem{lem}[thm]{Lemma}
\newtheorem{cor}{Corollary}[section]
\newtheorem{prop}[thm]{Proposition}

\theoremstyle{remark}
\newtheorem{Case}{\bf Case}[section]
\newtheorem{rem}[thm]{Remark}
\newtheorem{eg}{\sc Example}

\theoremstyle{definition}
\newtheorem{Def}{\bf Definition}[section]
\newtheorem*{pf}{\sc Proof}

\theoremstyle{definition}
\newtheorem{conj}{\bf Conjecture}

\newcommand{\N}{{\mathbb N}}
\newcommand{\ve}{{\varepsilon}}
\newcommand{\po}{{\partial}}

\newcommand{\del}{{\partial}}
\newcommand{\R}{{\mathbb{R}}}
\newcommand{\C}{{\mathbb C}}
\newcommand{\Z}{{\mathbb Z}}

\newcommand{\Rthree}{{\mathbb{R}^3}}
\newcommand{\Sph}{{\mathbb S}}

\newcommand{\sech}{{{\rm sech}}}
\newcommand{\tmu}{{ \tilde{\mu} }}

\begin{title}
{On Generalized Monopole Spherical Harmonics and the Wave Equation of
  a Charged Massive Kerr Black Hole}
\end{title}

\begin{author}
{Shabnam Beheshti\footnote{Department of Mathematics, Rutgers University, Piscataway, NJ.  beheshti@math.rutgers.edu} \qquad
Floyd L. Williams\footnote{Department of Mathematics, University of Massachusetts, Amherst, MA.  williams@math.umass.edu} }

\end{author}

\date{}

\maketitle

\begin{abstract}
We find linearly independent solutions of the Goncharov-Firsova
equation in the case of a massive complex scalar field on a Kerr black
hole.  The solutions generalize, in some sense, the classical monopole
spherical harmonic solutions previously studied in the massless cases.
\end{abstract}

\section{Introduction}
A \emph{spherical harmonic} $u(z)$ of one (real or complex)
variable $z$ can be defined as a solution of the linear differential
equation 
\begin{equation}\label{eq:sph}
(1-z^2)u^{\prime\prime}(z) -2zu^{\prime}(z)+\left[ \nu(\nu+1) -
    \frac{\mu ^2}{1-z^2}\right]u(z) = 0, 
\end{equation}
where $\nu$ and $\mu$ are fixed parameters.  A full discussion of this
equation can be found in Chapter 7 of ~\cite{Lebedev}, for example.
For fixed parameters $A, B, C$ and $a$, one can consider, more
generally, the linear differential equation 
\begin{equation}\label{eq:gensph}
(1-z^2)u^{\prime\prime}(z) -2zu^{\prime}(z)+\left[ -a +
    \frac{-Az^2+2Cz-B}{1-z^2}\right]u(z) = 0. 
\end{equation}
Although solutions of (\ref{eq:gensph}) are known, we present a brief
uniform approach based on the elegant theory of hypergeometric type
equations developed by A. Nikiforov and V. Uvarov ~\cite{Nikiforov}.
This approach allows us to construct, for example, explicit
generalized \emph{monopole spherical harmonics}, without the imposition of classical 
``quantization" (or integrality) conditions.  The classical theory is
developed in N. Vilenkin's book ~\cite{Vilenkin}, for example.  To
provide a context for these solutions we give a slightly more
general construction of the Wu-Yang quantized angular momentum operators ~\cite{WuYang}. 

In this paper we also find the general solution of the specific
differential equation 
\begin{eqnarray}\label{eq:main}
(1-z^2)u^{\prime\prime}(z) -2zu^{\prime}(z) + \left[ -(\lambda_l +
    \tmu ^2\alpha ^2) -2n\alpha kz \qquad \right. && \, \notag \\ 
\qquad \qquad \qquad \left. - \alpha ^2 (k^2-\tmu ^2)(1-z^2)  -
    \tfrac{m^2+n^2-2mnz}{1-z^2}\right]u(z) &=& 0, 
\end{eqnarray}
of Y. Goncharov, N. Firsova, equation (8) of ~\cite{GF} (where no
solutions are presented) which occurs in their work regarding topologically inequivalent configurations (TICs) of massive complex scalar fields on Kerr black holes.  Note that in
the special case when $\alpha=0$, (\ref{eq:main}) reduces to
(\ref{eq:gensph}) for appropriate choices of $A, B, C, a$.  We also
address the question of the orthogonality of our solutions to
(\ref{eq:main}) by transforming the equation to B\^ocher's equation
~\cite{MS}, and observing that the latter, in fact can be
written in Sturm-Liouville form.  Thus we prove the orthogonality
conjecture posed in ~\cite{GF}.  We remark that the parameter $\alpha$
relates to the black hole mass and angular momentum. 

The general program is the study of TICs of various fields on Kerr and other black holes, specifically in regards to their additional contributions to quantum effects.  These configurations, especially for non-zero $n$ in equation (\ref{eq:main}) of the paper, the so-called twisted case, at the physical level are linked with the presence of Dirac monopoles (given the quantization of charge condition) which are regarded as quantum objects residing in the black hole, and they are linked with the increase of Hawking radiation--a quantum effect.

\section{The Setting for Equation (\ref{eq:main})}
We provide in this section a brief sketch of how equation (\ref{eq:main}) 
arises and the meaning of the parameters $\lambda_l$, $\tmu$, $\alpha$, $m$, $n$, 
$k$.  The space-time vicinity of a rotating spherical object of mass $M$ and
 angular momentum $J$ is described by the Kerr metric $ds^2$ given as follows, where $r$, $\theta$, $\varphi$ are spherical coordinates (with $0 \leq \theta < \pi$ and $0 \leq \phi < 2 \pi$),
 $c$ is the speed of light, $G$ is Newton's gravitational constant, and 
$r_S\stackrel{def.}{=} 2GM/c^2$ is the Schwarzschild radius:
\begin{eqnarray}\label{eq:Kerr metric}
ds^2 &=& \left[1-\tfrac{r_S r}{\Sigma}\right]c^2 dt^2 -\tfrac{\Sigma}{\Delta}dr^2 -
\Sigma d\theta^2 \notag\\
&& \qquad - \left[r^2 + a^2 + \tfrac{r_S r a^2}{\Sigma}\sin^2 \theta \right]
\sin^2 \theta d\varphi^2 + \tfrac{2r_S r a \sin ^2 \theta}{\Sigma}c dt d\varphi ,
\end{eqnarray}
where $a \stackrel{def.}{=}J/Mc$ and 
$\Sigma \stackrel{def.}{=} r^2 + a ^2\cos^2\theta$, $\Delta \stackrel{def.}{=} r^2 -r_Sr + a^2$.
  This metric, which is an exact vacuum solution of Einstein's field equations,
 is used to describe a rotating black hole.  When $a=0$, that is $J=0$, the
 solution reduces to the Schwarzschild solution.  For convenience, we will choose $c=G=1$.

The authors Y. Goncharov, N. Firsova (G-F) have studied the contributions to
 Hawking radiation of TICs of complex scalar fields on several classes of black holes ~\cite{GF2, GF1}.  Such configurations can exist due
 to the non-triviality of the black hole topology $X=\mathbb{R}^2 \times \mathbb{S}^2$,
 and they correspond to smooth sections of complex line bundles $\mathcal{L}$ over $X$, which in turn
 are characterized by their Chern numbers $n \in \mathbb{Z}$ (for $\mathbb{Z}$ the set
 of integers).  This is the meaning of the $n$ in Equation (\ref{eq:main}).
 
$\mathcal{L}$ has curvature $F=dA$ (where $d$ denotes exterior differentiation) where in
 ~\cite{GF1} a gauge choice of the connection 1-form $A$ is given by
 \begin{equation}
 A = \frac{na \cos \theta}{e\Sigma}dt - \frac{n(r^2+a^2) \cos \theta}{e\Sigma} d\varphi ,
 \end{equation}
where the electric charge $e$ (the coupling constant) satisfies the Dirac quantization of charge
 condition
 \begin{equation}\label{eq:dirac}
 eq = n, \qquad 4 \pi q = \int_{S^2}F,
 \end{equation}
 $q$ being the magnetic charge.  By this gauge choice, the Maxwell equations $dF=0$, $d\ast F = 0$
 hold (where $\ast$ is the Hodge star operator), and moreover one has the Goncharov-Firsova 
 \emph{wave equation}
 \begin{equation}\label{eq:waveeqn}
 \Box \psi - \frac{1}{\Sigma \sin ^2 \theta}\left[ 2ni(a \sin ^2 \theta \tfrac{\del}{\del t} + 
\tfrac{\del}{\del \varphi} ) - n^2 \cos ^2 \theta \right]\psi = - \mu ^2 \psi
 \end{equation}
 where $\Box$ is the Laplace-Beltrami operator of the Kerr metric.  A 
separation of variables leads to a complete set of solutions in $L^2(X)$ of the following form
\begin{equation}\label{eq:fwml}
f_{\omega mln}(t,r,\theta ,\varphi) = \frac{e^{i\omega t}e^{-im\varphi}}{\sqrt{r^2+a^2}}S_l(\cos\theta, \omega , m,n)R(r,\omega ,m,l,n),
\end{equation}
with $m \in \mathbb{Z}$, $|m| \leq l$, $l = |n|,|n|+1, |n|+2, \ldots$; see page 1468 of ~\cite{GF1}.  
The radial functions $R(r,\omega ,m,l,n)$ are subject to a complicated second-order ordinary differential equation
 in the variable $r$, while the $S_l$ satisfy equation (\ref{eq:main}) for $z = \cos \theta$, $\alpha \stackrel{def.}{=} a/M$, 
$k\stackrel{def.}{=}\omega M$, $\tilde{\mu} \stackrel{def.}{=}\mu M$, and for suitable eigenvalues $\lambda_l$ indexed
 by $l$.  As stated in ~\cite{GF}, ~\cite{GoncharovYarevskaya}, solutions of the wave equation are necessary to compute contributions of TICs of complex scalar field on a black hole background; to calculate the expectation of the stress energy tensor, one requires the wavefunction explicitly; see ~\cite{GF}, page 1469.  Solutions of the wave equation govern perturbations of the Kerr metric (\ref{eq:Kerr metric}) (as is well known) such as those created, for example,when an orbiting body distorts the gravitational field of the black hole.

\section{Generalized monopole spherical harmonics}
We indicate how solutions $u(z)$ of equation (\ref{eq:gensph}) give rise to certain \emph{generalized monopole spherical harmonics}.  Without loss of generality, we may assume that $A=0$ in equation (\ref{eq:gensph}) by the following trivial argument.  Assume the equation
\begin{equation}\label{eq:A0}
(1-z^2)u^{\prime\prime}(z) -2zu^{\prime}(z)+\left[ -a +
    \frac{2Cz-B}{1-z^2}\right]u(z) = 0, \qquad \qquad |z|<1
\end{equation}
can be solved for any choice of parameters $a$, $B$, and $C$.  In turn, one may then solve (\ref{eq:gensph}) by replacing $a$ and $B$ in (\ref{eq:A0}) with $a-A$ and $B+A$, respectively, keeping $C$ the same in both equations. We associate to (\ref{eq:A0}) the parameter $a_0 = a_0(B,C)$ defined (up to a choice of sign) by
\begin{equation}\label{eq:a0}
a_0 ^2 \stackrel{def.}{=} \tfrac{1}{2}\left( B - \sqrt{B^2-4C^2} \right), \qquad |z|<1
\end{equation}
which will play a key role throughout this section.  Note that $B-a_0^2 = \tfrac{1}{2}\left(B + \sqrt{B^2-4C^2}\right)=C^2/a_0^2$,
 and therefore $B-2Cz+a(1-z^2)-(a+a_0^2)(1-z^2)=C^2/a_0^2-2Cz+a_0^2z^2 = (a_0z-C/a_0)^2$, which means that equation (\ref{eq:A0}) can be written as
 \begin{eqnarray}\label{eq:reduced sph}
 -(a+a_0^2)u(z) &=& -(1-z^2)u''(z) + 2zu'(z)   \\
&&  + \frac{1}{1-z^2}\left[B-2Cz+a(1-z^2)-(a+a_0^2)(1-z^2)\right]u(z) \nonumber \\
&=& -(1-z^2)u''(z)+2zu'(z)+\tfrac{1}{1-z^2}\left( a_0z-C/a_0\right)^2u(z). \nonumber
 \end{eqnarray}
That is, equation (\ref{eq:reduced sph}) compares exactly with the Wu-Yang equation (23) of ~\cite{WuYang} if the correspondence
$\Theta \leftrightarrow u$, $l(l+1) \leftrightarrow -a$, $q \leftrightarrow a_0$, $m \leftrightarrow -C/a_0$ between their notation and ours is made. Alternatively, we can define
\begin{equation}\label{eq:Theta}
\Theta(\theta) \stackrel{def.}{=} u(\cos \theta),
\end{equation}
in which case equation (\ref{eq:reduced sph}) assumes the form
\begin{equation}\label{eq:abstractwuyang}
-(a+a_0^2)\Theta = \left[ -\frac{1}{\sin \theta}\frac{\del}{\del \theta} \sin \theta \frac{\del}{\del \theta} + \frac{(a_0\cos \theta - C/a_0)^2}{\sin ^2 \theta}\right] \Theta,
\end{equation}
which is an abstract version of equation (22) of ~\cite{WuYang}.

We define abstract Wu-Yang quantized angular momentum operators
\begin{eqnarray}\label{eq:Lxyz}
\hat{L}_x &\stackrel{def.}{=}& i \sin \varphi \frac{\del}{\del \theta} + i \cos \varphi \cot \theta \frac{\del}{\del \varphi} - \frac{a_0 \sin \theta \cos \varphi}{1+\cos \theta}, \nonumber \\
\hat{L}_y &\stackrel{def.}{=}& -i \cos \varphi \frac{\del}{\del \theta} + i \sin \varphi \cot \theta \frac{\del}{\del \varphi} - \frac{a_0 \sin \theta \sin \varphi}{1+\cos \theta},  \\
\hat{L}_z &\stackrel{def.}{=}& -i \frac{\del}{\del \varphi} -a_0, \nonumber
\end{eqnarray}
in spherical coordinates $x=r \sin \theta \cos \varphi$, $y=r\sin \theta \sin \varphi$,  $z=r\cos \theta$.  Then for
\begin{equation}\label{eq:Lsquared}
\hat{L}^2 \stackrel{def.}{=} \hat{L}_x ^2 + \hat{L}_y ^2 + \hat{L}_z ^2
\end{equation}
it is possible to show that 
\begin{equation}\label{eq:Lsquared2}
\hat{L}^2 = -\frac{\del ^2}{\del \theta ^2} - \cot \theta \frac{\del}{\del \theta} - \frac{1}{\sin ^2 \theta}\frac{\del ^2}{\del \varphi ^2} + \frac{2i a_0}{1+\cos \theta}\frac{\del}{\del \varphi} + \frac{2a_0^2}{1+\cos \theta}.
\end{equation}

Analogous to the functions
\begin{equation}\label{eq:Zed}
Z_l ^m (\varphi , \theta) \stackrel{def.}{=} e^{im\varphi}P_l^{|m|}(\cos \theta)
\end{equation}
in the classical theory of spherical harmonics, with $l,m \in \mathbb{Z}$, $l \geq 0$, $|m| \leq l$, where $P_l^{|m|}$ is an associated Legendre function, we define the functions
\begin{equation}\label{eq:newZed}
Z(\varphi , \theta) \stackrel{def.}{=} e^{i (a_0 -C/a_0)\varphi} \Theta(\theta) \stackrel{def.}{=} e^{i (a_0 -C/a_0)\varphi} u(\cos \theta).
\end{equation}
Then, using equation (\ref{eq:abstractwuyang}) one can show that for $\hat{L}_z$ defined in (\ref{eq:Lxyz}) and for $\hat{L}^2$ computed in (\ref{eq:Lsquared2}) the following holds:
\begin{equation}\label{eq:LZeqns}
\hat{L}^2Z = -aZ , \qquad \hat{L}_z Z = -\frac{C}{a_0}Z.
\end{equation}
Given the above correspondence $\Theta \leftrightarrow u$, $l(l+1) \leftrightarrow -a$, $m \leftrightarrow -C/a_0$, the equations in (\ref{eq:LZeqns}) compare exactly with the formulas for the action of classical quantum mechanical angular momentum operators on hydrogenic wave functions, which is not surprising as the latter functions involve spherical harmonics.  See, for example, equations (5.4.5) and Theorem 5.2 of ~\cite{FWQM}; also compare equation (15) of ~\cite{WuYang}.  Thus, by way of definition (\ref{eq:a0}), a solution $u(z)$ of equation (\ref{eq:A0}) gives rise to an abstract monopole harmonic $Z(\varphi , \theta)$ in definition (\ref{eq:newZed}).  Of course, in ~\cite{WuYang} the monopole harmonics are properly viewed as fiber bundle sections.

We now outline a general technique to solve equation (\ref{eq:A0}) (equivalently (\ref{eq:main}) with $\alpha =0$), which we write as
\begin{equation}\label{eq:main2}
\sigma(z)^2 u''(z) + \sigma(z)\tilde{\tau}(z)u'(z)+\tilde{\sigma}(z)u(z) = 0 
\end{equation}
for $\sigma(z)\stackrel{def}{=}1-z^2$, $\tilde{\tau}(z)\stackrel{def}{=}-2z$, and 
$\tilde{\sigma}(z)\stackrel{def}{=}-\left[-2Cz+B+a(1-z^2)\right]$.

As $\sigma(z), \tilde{\sigma}(z), \tilde{\tau}(z)$ are polynomials with deg $\sigma(z),\tilde{\sigma}(z) \leq 2$ and deg $\tilde{\tau}(z) \leq 1$, we see that equation (\ref{eq:main2}) is in fact an equation of \emph{hypergeometric type} in the sense of Nikiforov and Uvarov ~\cite{Nikiforov}.
Thus, we can apply their elegant, universal technique, which provides in particular an associated, canonical quantization condition for equation (\ref{eq:A0})--without the usual recourse to power series methods.
The idea is to construct a canonical form of equation (\ref{eq:main2}), which is also of hypergeometric type and whose solutions relate to those of (\ref{eq:main2}).
For this, one proceeds as follows, where full details are also presented in Chapter 4 of ~\cite{FWQM}.  Here we consider the more general case with $C \neq 0$, for example.  In practice $C=mn$ for a ``magnetic quantum number" $m$ and Chern number $n$ of a black hole configuration, as in equation (\ref{eq:main}).  Given $\kappa \in \mathbb{C}$, define 
$f_{\kappa} = \tfrac{1}{4}(\tilde{\tau}-\sigma ')^2 +\kappa \sigma - \tilde{\sigma}$, which is a polynomial of degree $\leq 2$.  Assuming
 that the discriminant $\Delta(\kappa)$ of $f_\kappa$ vanishes, one can find a polynomial square root $p(f_\kappa)$.  In fact, we can choose $p(z)=a_0z-C/a_0$, for $a_0$ in definition (\ref{eq:a0}), where $C \neq 0 \Rightarrow a_0\neq 0$.  Then for $\pi_0 \stackrel{def}{=} \tfrac{1}{2}(\sigma ' - \tilde{\tau}) - p$, $\tau \stackrel{def}{=} \tilde{\tau} + 2\pi_0$ and $\lambda \stackrel{def}{=} \kappa + \pi_0 '$ ($\lambda \in \mathbb{C}$), it is shown in ~\cite{Nikiforov,FWQM} that if the functions $u(z)$, $y(z)$ are related by $u(z) = \Phi(z)y(z)$ for a function $\Phi(z)$ that satisfies $\Phi ' = \Phi \pi_0/\sigma$ (say $\Phi(z) = \mbox{exp} \int \pi_0(z)/\sigma(z) dz$), then $u(z)$ solves equation (\ref{eq:main2}) (equivalently equation (\ref{eq:A0})) if and only if $y(z)$ solves the simpler, canonical equation
\begin{equation}\label{eq:gen canonical form}
\sigma(z)y ''(z) + \tau(z) y '(z) + \lambda y(z) = 0,
\end{equation}
which in our case is 
\begin{equation}\label{eq:canonical form}
(1-z^2)y''(z) + \left[ (-2-2a_0)z+\tfrac{2C}{a_0}\right]y'(z) + \left(-a_0^2-a-a_0 \right) y(z) = 0.
\end{equation}

On the other hand, under the change of variables $v(z) = y(-1+2z)$, equivalently $y(z) = v(\tfrac{z+1}{2})$, equation (\ref{eq:canonical form}) is transformed to the classical Gauss Hypergeometric Equation
\begin{equation}\label{eq:gauss}
z(1-z)v''(z) + \left[ \,\underline{\gamma} - (\underline{\alpha} + \underline{\beta} + 1)z \right]v'(z) -\underline{\alpha}\underline{\beta} v(z)=0,
\end{equation}
with $\underline{\alpha} = \tfrac{1}{2}\left[ 1+2a_0 + \sqrt{1-4a}\right]$, 
$\underline{\beta} =\tfrac{1}{2}\left[ 1+2a_0 - \sqrt{1-4a}\right]$, and $\underline{\gamma} = 1+a_0 + \tfrac{C}{a_0}$.  A solution of equation (\ref{eq:gauss}) is $v(z) = F(\underline{\alpha} , \underline{\beta} ; \underline{\gamma} ; z)$, of course, where $F$ is the Gauss hypergeometric
 function.  Then $y(z) = v(\tfrac{z+1}{2})=F(\underline{\alpha} , \underline{\beta} ; \underline{\gamma} ; \tfrac{z+1}{2})$ solves equation (\ref{eq:canonical form}).  

Also, we can take $\Phi(z)=(1-z)^{\alpha/2}(1+z)^{\beta/2}$ on $|z|<1$, for $\alpha \stackrel{def}{=} a_0 - C/a_0$, $\beta \stackrel{def}{=} a_0 + C/a_0$ and thus obtain the solution
\begin{equation}\label{eq:solutionu}
u(z) = \Phi(z)y(z) = (1-z)^{\alpha/2}(1+z)^{\beta/2}F(\underline{\alpha} , \underline{\beta} ; \underline{\gamma} ; \tfrac{z+1}{2})
\end{equation}
of equation (\ref{eq:A0}) on $|z|<1$, among other possible solutions.  Notice that this construction always gives rise to single-valued waves, since we consider the restriction $|z|<1$, a simply connected domain where the logarithm is defined.  The corresponding generalized monopole spherical harmonics $Z(\varphi, \theta)$ in definition (\ref{eq:newZed}) is given by
\begin{equation}\label{eq:newZed2}
Z(\varphi , \theta) = e^{i \alpha \varphi}(1-\cos \theta)^{\alpha /2}(1+\cos \theta)^{\beta / 2} F\left( \underline{\alpha} , \underline{\beta} ; \underline{\gamma} ; \frac{1+ \cos \theta}{2}\right),
\end{equation}
where $\underline{\alpha} , \underline{\beta}, \underline{\gamma}$ are defined as previously.

Equation (\ref{eq:gen canonical form}) (for general polynomials $\sigma(z) , \tau(z)$ and scalar $\lambda$ with deg $\sigma(z)\leq 2$ and deg $\tau(z) \leq 1$) admits polynomial solutions $y_N(z)$ of degree $\leq N$ provided the \emph{quantization} or \emph{integrality}
   condition $\lambda = \lambda_N \stackrel{def}{=} -N\tau ' - \tfrac{N(N+1)}{2}\sigma ''$, $N=0,1,2, \ldots$, is satisfied. In the present situation with equation (\ref{eq:canonical form}), this condition is satisfied specifically when $-a = (a_0 + N)(a_0 + N + 1)$, in which case $y_N(z)$ is the Jacobi Polynomial $y_N(z)= P_N^{(\alpha, \beta)}(z)$ (with $\alpha \stackrel{def}{=} a_0 - C/a_0, \beta \stackrel{def}{=} a_0 + C/a_0$, as above), $u(z)=u_N(z)=(1-z)^{\alpha/2}(1+z)^{\beta/2}P_N^{(\alpha, \beta)}(z)$ on $|z|<1$, and
\begin{equation}\label{eq:finallabel}
Z(\varphi , \theta) = e^{i\alpha \varphi}(1-\cos \theta)^{\alpha /2}(1+\cos \theta)^{\beta /2}  P_N^{(\alpha , \beta)}(\cos \theta).
\end{equation}

\section{Solutions of the Goncharov-Firsova equation}
At this point we turn to solving the Goncharov-Firsova equation (\ref{eq:main})
\begin{equation}\label{eq:transf-main}
S_{l}''(z) - \tfrac{2z}{1-z^2}S_l '(z) - \left[\tfrac{\lambda_l+\tilde{\mu}^2\alpha^2}{1-z^2}+\alpha^2(k^2-\tilde{\mu}^2) + \tfrac{2n\alpha kz}{1-z^2} + \tfrac{m^2+n^2-2mnz}{(1-z^2)^2}\right]S_l(z) = 0,
\end{equation}
for the spherical component of the solution $f_{\omega m l n}$ to equation (\ref{eq:fwml}) on $|z|<1$.  Equation (\ref{eq:transf-main}) assumes a \emph{B\^ocher form} ~\cite{MS}:
\begin{equation}\label{eq:Bocher}
S_l''(z)+P(z)S_l'(z)+Q(z)S_l(z)= 0, \qquad \qquad |z|<1,
\end{equation}
where $P(z)= \frac{1}{2}\left[\frac{m_1}{z-a_1}+\frac{m_2}{z-a_2}\right]$, $Q(z)=\frac{1}{4}\left[\frac{Q_0+Q_1z+Q_2z^2+Q_3z^3+Q_4z^4}{(z-a_1)^{m_1}(z-a_2)^{m_2}}\right]$, for $m_1=m_2=2$, $a_1=-1$, $a_2=1$, and for $Q_j$ defined as
\begin{eqnarray}\label{eq:q-eqns}
Q_0 &=& -4\left[\alpha^2k^2 + \lambda_l + m^2+n^2\right] \nonumber\\
Q_1 &=& 8n(m-\alpha k) \nonumber\\
Q_2 &=& 4\left[2\alpha^2k^2 - \alpha ^2\tilde{\mu}^2 + \lambda_l\right]\\
Q_3 &=& 8n\alpha k \nonumber\\
Q_4 &=& -4\alpha^2 (k^2-\tilde{\mu}^2). \nonumber
\end{eqnarray}
B\^ocher's equation has a natural suitableness for the application of the Frobenius method. The functions $zP(z)$ and $z^2Q(z)$ are analytic on $|z|<1$ and in particular at $z_0=0$.  Consider the power series expansions $zP(z)=\sum_{n=0}^\infty A_n z^n$, $z^2Q(z) = \sum_{n=0}^\infty B_n z^n$ and $S_l(z) = z^\beta \sum_{n=0}^\infty C_nz^n$, about the regular singular point $z_0=0$, we have $\beta$ satisfying the indicial equation $\beta^2 + (A_0 -1)\beta + B_0 = 0$.  We see that $A_0 = B_0 =0$ so that $\beta = 0$ or $1$, in fact.  By the general theory, the recurrence formula for $C_n$
\begin{equation}\label{eq:recurrence}
C_n = C_n(\beta) = \frac{-1}{(\beta + n)(\beta + n -1)}\sum_{k=0}^{n-1} \left[(\beta+k)A_{n-k} + B_{n-k}\right]C_k,
\end{equation}
allows us to compute the $A_n$, $B_n$, $n \geq 0$.  Taking $C_0=1$, we find that $A_0 = A_1 = A_3 = \cdots = A_{2t-1}=0$, $A_2=A_4=\cdots = A_{2t}=-2$, $B_0=B_1=0$, $B_2=Q_0/4$, $B_3=Q_1/4$, $B_4=(2Q_0+Q_2)/4$, $B_5=(2Q_1+Q_3)/4$ and 
\begin{equation}
B_j = \left\{ \begin{array}{ll}
\frac{1}{4}(\frac{j}{2}Q_0+\frac{j-2}{2}Q_2+\frac{j-4}{2}Q_4) & \textrm{for } j\geq 6, \,j\,\textrm{ even}\\
\,&\,\\
\frac{1}{4}(\frac{j-1}{2}Q_1+\frac{j-3}{2}Q_3)                & \textrm{for } j\geq 6, \,j\,\textrm{ odd}.
\end{array}
\right.
\end{equation}
Thus, for the larger root $\beta =1$ of the indicial equation, we have expressed a solution $S_{l,1}(z)$ of equation (\ref{eq:transf-main}) on $|z|<1$ in terms of the parameters $\alpha$, $k$, $\lambda_l$, $m$, $n$, $\tilde{\mu}$ by the coefficients $C_j(1)$, recursively defined.

To obtain a second solution, linearly independent from the first, we use the smaller root of the indicial equation, namely $\beta =0$.  Then $S_{l,2}(z) \stackrel{def.}{=} d_0S_{l,1}(z)\log z + z^\beta \sum_{n=0}^\infty D_n z^n$.  Recalling that $d_0= \lim _{\beta \rightarrow 0} (\beta -0)C_N(\beta)$, where $N=$larger root - smaller root $=1$, we find that $d_0=0$ so that $S_{l,2}(z) = \sum_{n=0}^\infty D_n z^n$, with $D_n$ given by $D_0=C_0$, $D_n = \frac{d}{d\beta}(\beta -0)C_n(\beta) |_{\beta =0}=C_n(0)$ for $n\geq 1$.  The coefficients  $C_j,D_j$, $0\leq j \leq 8$, given in the tables, are expressed in terms of $Q_j$  from (\ref{eq:q-eqns}), which are in turn, in terms of the physical data.

\begin{table}[ht]
\centering
\begin{tabular}{c|l}
$j$ & $C_j(1)$ \\
\hline
0  & 1 \\[1ex]
1  & 0 \\[1ex]
2  & $\tfrac{1}{3} - \tfrac{Q_0}{24}$ \\[1ex]
3  & $\tfrac{-Q_1}{48}$ \\[1ex]
4  & $\tfrac{1}{5} - \tfrac{Q_0}{24} + \tfrac{Q_0^2}{1920} - \tfrac{Q_2}{80}$ \\[1ex]
5  & $\tfrac{-Q_1}{40} + \tfrac{Q_0 Q_1}{1920} - \tfrac{Q_3}{120}$ \\[1ex]
6  & $\tfrac{1}{7} - \tfrac{7 Q_0}{180} + \tfrac{Q_0^2}{1152} - \tfrac{Q_0^3}{322560} + \tfrac{Q_1^2}{8064} - \tfrac{17 Q_2}{1008} + \tfrac{13 Q_0 Q_2}{40320} - \tfrac{Q_4}{168}$ \\[1ex]
7  & $\tfrac{-43 Q_1}{1680} + \tfrac{13 Q_0 Q_1}{13440} - \tfrac{Q_0^2 Q_1}{215040} + \tfrac{Q_1 Q_2}{6720} - \tfrac{41 Q_3}{3360} + \tfrac{Q_0 Q_3}{4480}$ \\[1ex]
8  & $\tfrac{1}{9} - \tfrac{409 Q_0}{11340} + \tfrac{19 Q_0^2}{17280} - \tfrac{Q_0^3}{138240} + \tfrac{Q_0^4}{92897280} + \tfrac{53 Q_1^2}{207360} - \tfrac{13 Q_0 Q_1^2}{5806080}$ \\[1ex]
$\,$ & $- \tfrac{239 Q_2}{12960} + \tfrac{233 Q_0 Q_2}{362880} - \tfrac{17 Q_0^2 Q_2}{5806080} + \tfrac{Q_2^2}{23040} + \tfrac{7 Q_1 Q_3}{69120} - \tfrac{Q_4}{108} + \tfrac{Q_0 Q_4}{6048}$
\end{tabular}
\label{table:coeffs1}
\caption{Coefficients $C_j$ of series solutions to Equation (\ref{eq:transf-main})}
\end{table}

\begin{table}[ht]
\centering
\begin{tabular}{c|l}
$j$ & $D_j=C_j(0)$ \\
\hline
0  & 1 \\[1ex]
1  & 0 \\[1ex]
2  & $\tfrac{-Q_0}{8}$ \\[1ex]
3  & $\tfrac{-Q_1}{24}$ \\[1ex]
4  & $\tfrac{-Q_0}{12} + \tfrac{Q_0^2}{384} - \tfrac{Q_2}{48}$ \\[1ex]
5  & $\tfrac{-3Q_1}{80} + \tfrac{Q_0 Q_1}{480} - \tfrac{Q_3}{80}$ \\[1ex]
6  & $\tfrac{-23 Q_0}{360} + \tfrac{Q_0^2}{288} - \tfrac{Q_0^3}{46080} + \tfrac{Q_1^2}{2880} - \tfrac{Q_2}{45} + \tfrac{7 Q_0 Q_2}{5760} - \tfrac{Q_4}{120}$ \\[1ex]
7  & $\tfrac{-11 Q_1}{336} + \tfrac{43 Q_0 Q_1}{13440} - \tfrac{Q_0^2 Q_1}{35840} + \tfrac{Q_1 Q_2}{2688} - \tfrac{5 Q_3}{336} + \tfrac{11 Q_0 Q_3}{13440}$ \\[1ex]
8  & $\tfrac{-11 Q_0}{210} + \tfrac{11 Q_0^2}{2880} - \tfrac{Q_0^3}{23040} + \tfrac{Q_0^4}{10321920} + \tfrac{11 Q_1^2}{17920} - \tfrac{Q_0 Q_1^2}{92160}$\\[1ex]
 & $\tfrac{-71 Q_2}{3360} + \tfrac{41 Q_0 Q_2}{20160} - \tfrac{11 Q_0^2 Q_2}{645120} + \tfrac{Q_2^2}{10752} + \tfrac{13 Q_1 Q_3}{53760} - \tfrac{3 Q_4}{280} + \tfrac{Q_0 Q_4}{1680}$ 
\end{tabular}
\label{table:coeffs2}
\caption{Coefficients $D_j$ of series solutions to equation (\ref{eq:transf-main})}
\end{table}
In summary, we produce power series solutions to Equation (\ref{eq:transf-main}) having coefficients $C_j(\beta)$, $\beta = 0,1$ encoding the physical data via (\ref{eq:q-eqns}).  Notice that the series does not truncate in general, as $\alpha$ is nonzero.  However, in the case when $\alpha =0$, (\ref{eq:main}) reduces to (\ref{eq:gensph}) and the solutions are hypergeometric functions; in certain instances, such functions reduce further to Jacobi, Laguerre, or Legendre polynomials, among others.

It is interesting to note that the B\^ocher equation derived appears in the study of the Willmore functional, or extrinsic Polyakov action.  Considering a reduction of the Weierstrass formula for surfaces in $\mathbb{R}^3$, Konopelchenko and Taimanov examine the system $r'(x) = -r(x)/2 + 2p(x)s(x)$, $s'(x) = s(x)/2 - 2p(x)r(x)$ in equation (7) of ~\cite{Konopelchenko}.  By choosing potential $p(x)= (x-a_1)^{-1}(x-a_2)^{-1}$, the equation in $r(x)$ can be re-expressed as a  B\^ocher equation.  Note $p(x)$ assumes the same form as $P(z)$ appearing in \eqref{eq:Bocher}; in this case it is not necessary to assume $m_1=m_2=2$, and the expression for $Q(z)$ becomes a polynomial of degree $m_1+m_2$.   One may still view this as an equation of Bo\^cher type and proceed as before ~\cite{MoonSpencer}.

\section{Orthogonality of solutions}
We may also write equation (\ref{eq:main}), equivalently equations (\ref{eq:transf-main}), (\ref{eq:Bocher}) in Sturm-Liouville form, from which orthogonality of solutions follows. Setting $U(x) = x^2-1$, $-1 < x < 1$, we observe that 
\begin{equation}
\frac{1}{U(x)}\frac{d}{dx}\left[U(x) S_l'(x)\right] = S_l ''(x) - \frac{2x}{1-x^2}S_l '(x),
\end{equation}
which means that we can write equation (\ref{eq:transf-main}) in Sturm-Liouville form
\begin{equation}\label{eq:Sturm-Liouville}
\frac{d}{dx}\left[U(x) S_l'(x)\right] + \left[ V(x) + \lambda_l + \tilde{\mu}^2\alpha ^2 \right]S_l(x) = 0
\end{equation}
for $V(x) \stackrel{def.}{=} \alpha^2(k^2-\tilde{\mu}^2)(1-x^2) + 2n\alpha k x + \frac{m^2+n^2-2mnx}{1-x^2}$. By general principles, one has orthogonality of solutions.  Namely, we have if $\lambda_l \neq \lambda_m$, then $\int_{-1}^1 S_l(x)S_m(x) dx =0$, as conjectured in ~\cite{GF}.


\section{Conclusions}  
Apart from the presentation of a generalized construction of quantized Wu-Yang angular momentun operators, we provide further knowledge regarding the solutions of the Firsova-Goncharov wave equation (\ref{eq:waveeqn}).  

One cannot calculate any of the quantum effects, touched on lightly in the introduction, without information on solutions $\psi$ of the wave equation, even in the simpler massless and untwisted cases, with $\mu$ in (\ref{eq:dirac}) and  n in (\ref{eq:main})  both equal to zero.  For example, $\psi$ and the metric (\ref{eq:Kerr metric}) determine the energy momentum tensor, whose vacuum expectation value, in turn, determines the luminosity of the Hawking radiation.  On the other hand, due to the complexity of various formulas involved, it is necessary to further develop numerical schemes to facilitate the computations; a numeric algorithm to approximate the eigenvalues $\lambda_l$  of equation (\ref{eq:main}) is yet in the works.  Hopefully, progress on this front as well as other aspects of the general program will be made in future work.


\bibliography{monopole}
\bibliographystyle{plain}

%
%
%
%
%
%
%
%
%

\end{document}